\def   \ni {\noindent}

\def   \ssk {\vskip  5truept}

\def   \bsk {\vskip 15truept}

\def   \newline {\hfil\break}

\documentstyle[epsfig]{article}
\begin{document}

\hsize 5truein
\vsize 8truein
\font\abstract=cmr8
\font\keywords=cmr8
\font\caption=cmr8
\font\references=cmr8
\font\text=cmr10
\font\affiliation=cmssi10
\font\author=cmss10
\font\mc=cmss8
\font\title=cmssbx10 scaled\magstep2
\font\alcit=cmti7 scaled\magstephalf
\font\alcin=cmr6 
\font\ita=cmti8
\font\mma=cmr8
\def\ref{\par\noindent\hangindent 15pt}
\null


\setlength{\unitlength}{1mm}
\def\fwb{58mm}
\def\fwc{50mm}
\newcommand{\gray}{$\gamma$-ray\ }
\newcommand{\la}{\le}
\newcommand{\ga}{\ge}
\hyphenation{brems-strahl-ung}

{\footnotesize \it \vspace{-14\baselineskip} \noindent 
   Proc.\ 3rd INTEGRAL Workshop ``The Extreme Universe'',
   14--18 Sep.\ 1998, Taormina, Italy
   \\ \rule[3ex]{124mm}{0.1mm}
\vspace{1ex} \vspace{11\baselineskip} }


\title{\ni A PAIR PLASMA MODEL FOR PKS 0208--512}                                               

\bsk \bsk
\author{\ni I.V. Moskalenko$^{1,2}$, W.Collmar$^1$}                                                       
\bsk
\affiliation{1) Max-Planck-Institut f\"ur extraterrestrische Physik, 
D-85740 Garching, Germany}                                                

\affiliation{2) Institute for Nuclear Physics, Moscow State University, 
119 899 Moscow, Russia}                                                
\bsk
\baselineskip = 12pt

\abstract{ABSTRACT \ni
Assuming that the enhanced MeV emission from the MeV blazar PKS
0208--512 observed by COMPTEL on occasions is due to annihilation in a
pair plasma, we estimate parameters of the annihilation region: its
size, the number density of particles, and the variability timescale.
The values we find are in accord with our present-day knowledge of
emission mechanisms operating in AGNs. The constructed model implies
that the blueshifted annihilation emission is an intrinsic property of
all AGN jet models, and predicts an anticorrelation between the
high-energy emission and the annihilation flux.
}                                                    
\bsk
\baselineskip = 12pt

\bsk
\baselineskip = 12pt


\text{\ni 1. INTRODUCTION
\ssk
\ni     

During six years of operation about 70 blazars have been detected by
the EGRET telescope at $\ge$100 MeV (e.g., Hartman et al.\ 1997).
Although the origin of the blazar \gray emission is still discussed,
the inverse Compton scattering of soft photons off relativistic
electrons in a collimated jet is widely accepted (e.g., Dermer et
al.\ 1992, Maraschi et al.\ 1992).  Other models associate the
$\gamma$-ray emission with $\pi^0$-production by relativistic nucleons
(e.g., Mastichiadis \& Protheroe 1990, Mannheim 1993).

COMPTEL observations indicate a class of \gray blazars which display
spectral energy distributions peaking in a narrow energy band
at a few MeV (Bloemen et al.\ 1995, Blom et al.\ 1995, 1996).  The
blazar PKS 0208--512 was detected in the analysis of COMPTEL data by Blom
et al. (1995). The signal was obtained by combining data from May
8--13, 1993 and June 3--14, 1993, yielding a strong flux in the 1--3 MeV band,
and only upper limits at lower and higher COMPTEL energies.
Contemporary observations with EGRET yielded a significant detection above 100
MeV.  The peak in the $\nu F_\nu$ spectrum of PKS 0208--512 occurs at
MeV energies (Kanbach 1996).
Remarkable is the anticorrelation in the source flux at COMPTEL and
EGRET energy ranges for the CGRO Phases I and II.

The MeV features observed by COMPTEL can not be explained in the
framework of jet-scattered background emission and  therefore probably have a 
different origin.  The most natural one seems the interpretation of Doppler
boosted $e^+e^-$-annihilation radiation (e.g., Henri et al.\ 1993,
Roland \& Hermsen 1995).  The clear and pronounced feature observed from
PKS 0208--512 allows model fitting accurate enough to derive such
parameters as the pair plasma temperature, the bulk Lorentz factor,
viewing angle, and the intrinsic annihilation luminosity (Skibo et al.
1997); it has also been shown that the existence of a class of MeV
blazars can be understood by orientation effects, and is consistent
with AGN unification scenarios.

\begin{figure}[tb]
   \begin{picture}(120,41)(0,0)
      \put(0,0){%
         \makebox(60,0)[lb]{\psfig{file=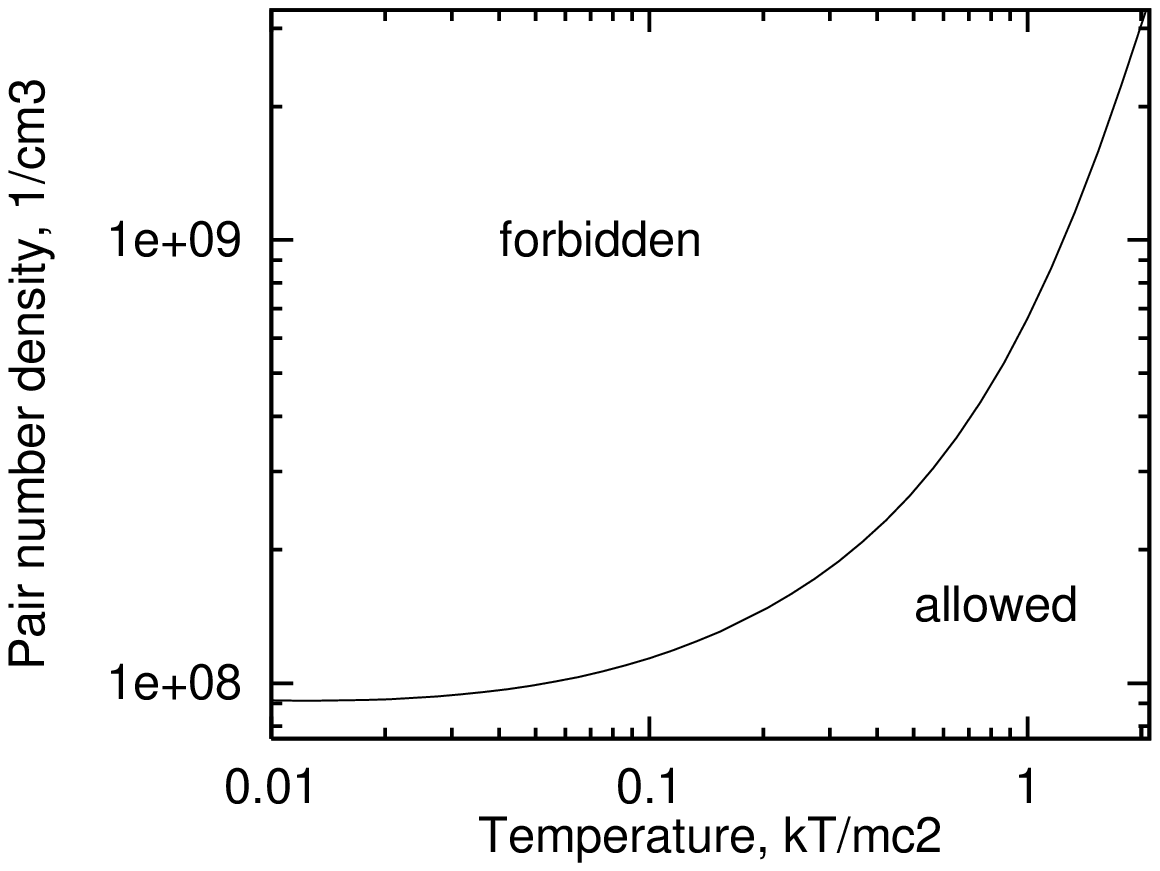,width=\fwb,clip=}}}
      \put(65,0){%
         \makebox(60,0)[lb]{\psfig{file=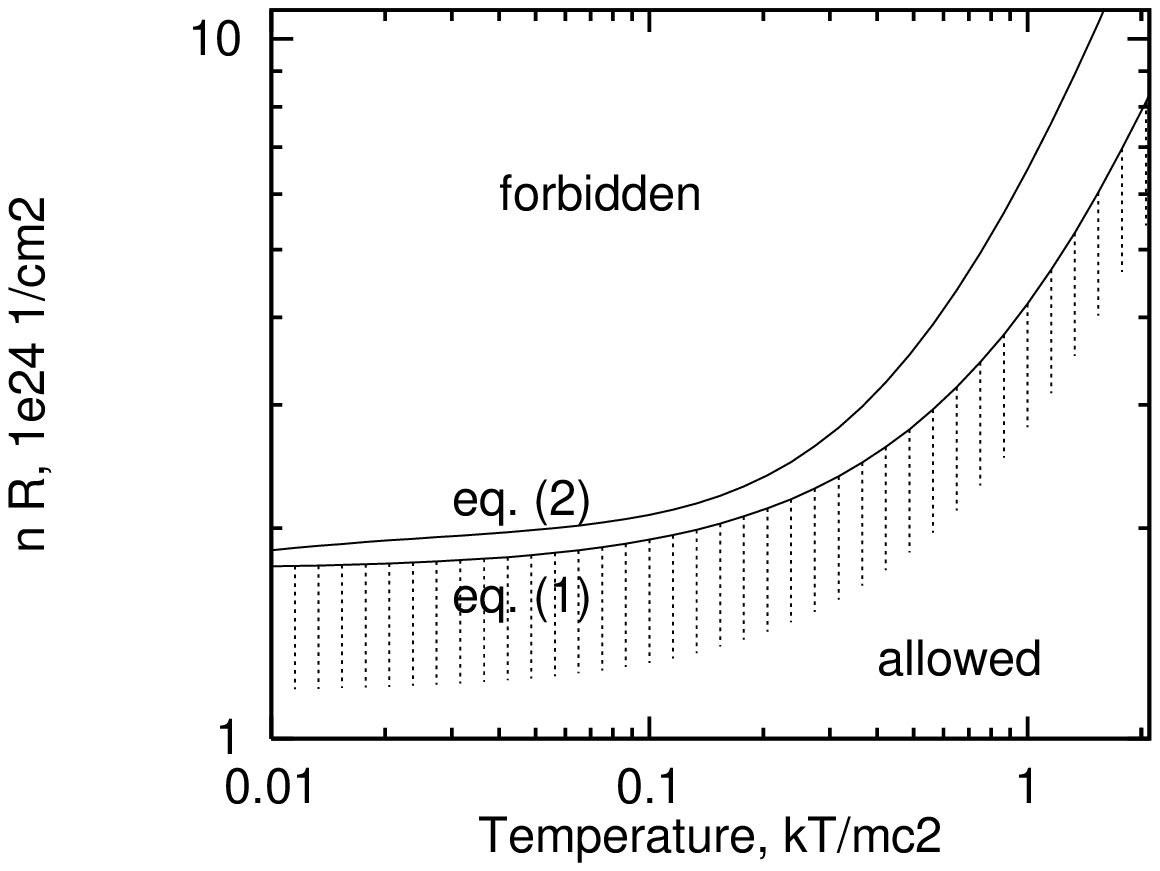,width=\fwb,clip=}}}
   \end{picture}
\caption{ FIGURE 1.
Left:
The upper limit for pair number density (eq.~[1]), 
$L_a^\ast = 10^{46}$ erg s$^{-1}$.
Right:
The upper limits for the product $(n_\pm^\ast R^\ast)$ vs.\ plasma
temperature.
\label{fig1} }
\end{figure}



\bsk
\ni 2. ANNIHILATION OF THE THERMAL PAIR PLASMA 
\ssk
\ni 

A distinct annihilation line from a pair plasma can only be observed
if the plasma is optically thin.  The opacity
of an optically thin pair plasma is dominated by Compton scatterings.
We thus consider only the case when the Compton scattering
optical depth is less then unity
$\tau_C(\theta)= 2 n_\pm^\ast R^\ast \sigma_{KN}(\theta) \la 1$,
where $n_\pm^\ast$ is the number density of pairs, $R^\ast$ is the
radius of the plasma blob, $\sigma_{KN}(\theta)\approx
\sigma_{KN}(\theta;\epsilon_{\max})$ is the Klein-Nishina cross section
averaged over the thermal electron distribution, $\theta=kT/mc^2$ is
the plasma temperature, and $\epsilon_{\max}$ is the energy of the
center of the annihilation line.  An asterisk marks variables in the
comoving frame.

The annihilation luminosity can be obtained as
$L_a^\ast = 
   4\pi c R^{\ast 2} t_{esc} \langle\dot{n}\epsilon\rangle$,
where $t_{esc}\sim R^\ast/c$ is the photon escape time in the optically
thin plasma, and the annihilation emissivity is given by
$\langle\dot{n}\epsilon\rangle\approx 
   2 \pi r_e^2 c n_\pm^{\ast 2} A(\theta)\, \bar{\epsilon}$.
Here the factor of 2 accounts for 2 photons per annihilation event, $r_e$
is the classical electron radius, $\bar{\epsilon} \sim \epsilon_{\max}$
is the average energy of the annihilation photon, $A(\theta)$ is
the dimensionless annihilation rate (Moskalenko \& Jourdain 1997)
averaged over the thermal pair distribution, and
$t_a^\ast = [\pi r_e^2 c n_\pm^\ast A(\theta)]^{-1}$
is the intrinsic annihilation timescale.

The condition $\tau_C\le 1$ yields estimates for the pair
density and the blob radius
\begin{equation}
\label{eq.10}
n_\pm^\ast\la \pi^2 r_e^2 c A(\theta)\, \bar{\epsilon}/
   [\sigma_{KN}^3(\theta) L_a^\ast], \qquad
n_\pm^\ast R^\ast\la [2\sigma_{KN}(\theta)]^{-1},
\end{equation}
which depend only on the luminosity $L_a^\ast$ and the plasma temperature.
Converting $t_a^\ast$ into the blob size by using the light
travel time $2 R^\ast \la ct_a^\ast$ yields another estimate, 
\begin{equation}
\label{eq.11}
n_\pm^\ast R^\ast \la [2 \pi r_e^2 A(\theta)]^{-1}.
\end{equation}
The upper limits for $n_\pm^\ast$ and $(n_\pm^\ast R^\ast)$ are shown
in Fig.~1. Expr.\ (1)--(2) agree better than a factor
of 2, showing that the variability timescale in the comoving frame
is approximately twice the intrinsic annihilation timescale,
$\Delta_d^\ast\sim 2t_a^\ast$.


These formulae allow the derivation of a lower limit on the observed
variability timescale of the annihilation line from a single
pair plasma blob ($\tau_C \le 1$):
\begin{equation}
\label{eq.12}
\Delta_d = 
   \frac{1+z}{{\cal D}}\, \frac{3\, t_a^\ast}{2} \ga 
   \frac{1+z} {\cal D}\,
   \frac{ 3 \sigma_{KN}^3(\theta)\, L_a^\ast }
   { 2 \pi \bar{\epsilon}\, [\pi r_e^2 c\, A(\theta)]^2}
   ,
\end{equation}
where $z$ is the redshift, and $\cal D$ is the Doppler factor, and we took
also into account that the variation timescale in $L^\ast_a$
should be a factor of 2 shorter then $t_a^\ast$
($\dot{L}_a/L_a = 2 \dot{n}_\pm/n_\pm$).  The intrinsic
annihilation timescale, $t_a^\ast(\theta)$, is shown in
Fig.~2.

Since the annihilation line from the pair plasma blob is most effectively
emitted when $\tau \sim 1$, the pair number density is most probably
equal to its upper limit (eq.~[1]). Thus the blob radius is $R^\ast
\sim [2 n_\pm^\ast \sigma_{KN}(\theta)]^{-1}$, and the variability
timescale should be close to its lower limit (eq.~[3]).


\begin{figure}[tb]
   \begin{picture}(120,41)(0,0)
      \put(0,0){%
         \makebox(60,0)[lb]{\psfig{file=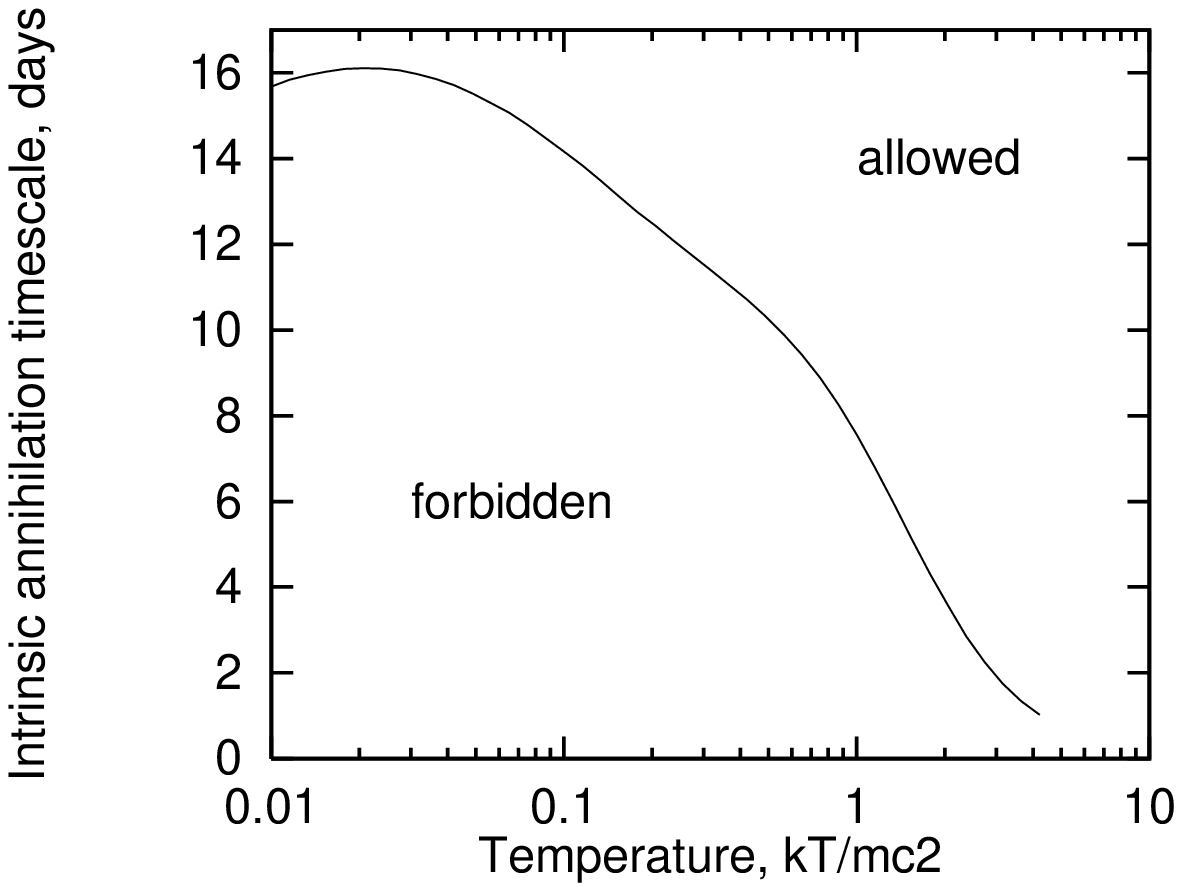,width=\fwb,clip=}}}
      \put(67,-4){%
         \makebox(60,0)[lb]{\psfig{file=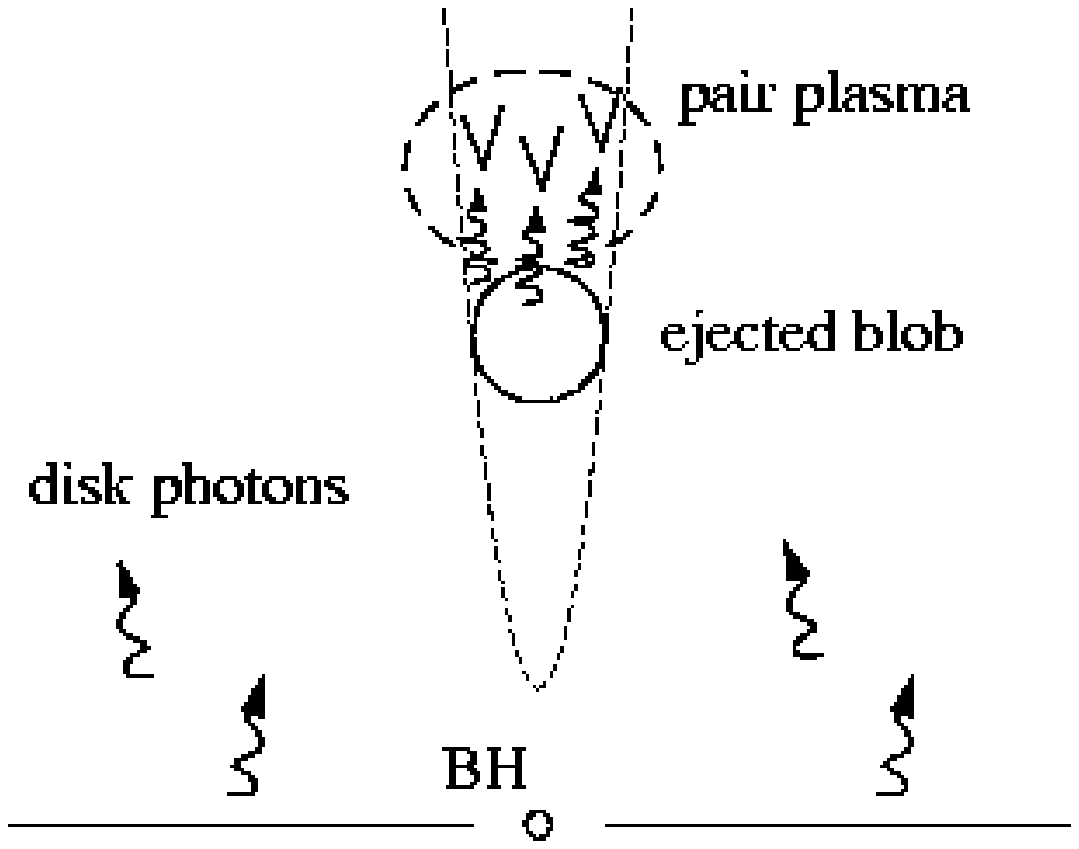,width=\fwc,clip=}}}
   \end{picture}

\parbox[t]{58mm}{%
\caption{ FIGURE 2.
The intrinsic annihilation timescale, $t_a^\ast(\theta)$,
 for $L_a^\ast = 10^{46}$ erg s$^{-1}$.
\label{fig2} }
}\hspace{4mm}
\parbox[t]{58mm}{%
\caption{ FIGURE 3.
Schematic view illustrating the model.
\label{fig3} }
}
\end{figure}


Assuming that the enhanced MeV emission observed from PKS 0208--512 is due to
annihilation in the outflowing thermal pair plasma blob, we adopt the
parameters obtained by Skibo et al.\ (1997) from the spectral fitting:
$L_a^\ast =3.3 \times10^{46}$ erg s$^{-1}$, $\theta=0.5$, a bulk
Lorentz factor $\Gamma=3$, and a viewing angle $\psi=18^\circ$.

Eq.~(1) allows to put the upper limit on the pair number density of the
plasma blob as $n_\pm^\ast\la 8.2\times10^7$ cm$^{-3}$ (Fig.~1 left).
The obtained variability timescale, $\Delta_d \ga 31.5$ days ($z=1$,
${\cal D}\approx 3.2$), agrees well with the total duration of 38 days
(VPs 220.0, 224.0, + time in between) for which COMPTEL observed this
MeV feature from PKS 0208--512.  Using the upper limit, $n_\pm^\ast\sim
8.2\times10^7$ cm$^{-3}$, provides an estimate for the radius of
$R^\ast\sim 3.4\times 10^{16}$ cm assuming the best observability
conditions (see Fig.~1 right).  This is a typical estimate for the blob
radius (Dermer et al.\ 1997).  The blob size $d^\ast\sim c
\Delta_0 {\cal D}/(1+z)\approx 3.34\times 10^{16}$ cm inferred from the
shortest variability timescale of PKS 0208--512 observed by EGRET
($\Delta_0\approx 8$ days, von Montigny et al.\ 1995) matches well the
above estimate for $R^\ast$.  The annihilation timescale (eq.~[3])
provides us also with a lower limit for the high energy emission site,
which due to the $\gamma\gamma$-transparency arguments should be
located at the distance of
$z_i\ga 
\Gamma c \frac{t_a^\ast}{2} \approx 
0.01 {\rm\ pc\ } \left(\Gamma/3\right)
   \left(L_a^\ast/3.3 \times10^{46}{\rm\ erg\ s}^{-1}\right)$
from the central engine.

\bsk
\ni
3. THE MODEL
\ssk
\ni

The $\gamma\gamma$-opacity effects play an important role when
considering the generation of high energy $\gamma$-ray emission.  In the
early stage of the flare when the ejected plasma blob is still close to
the central engine, it is embedded in a dense background radiation
field.  A huge optical depth due to the $\gamma\gamma$-pair production
prevents the high energy photons from escaping. The pairs appear
preferrentially ahead of the ejected blob and move with relativistic
speed in the same direction as follows from the energy-momentum
conservation law (Fig.~3).  A relatively small Lorentz factor of the pairs
and large optical depth of the pair plasma itself makes it easier to
establish (quasi-) thermal equilibrium.  The annihilation of
pairs gives rise to annihilation photons, for which the background
radiation field is transparent. The annihilation line becomes visible
as soon as the optical depth of the pair plasma cloud becomes less then
unity while its size is large enough to produce the observed
flux.  The high energy \gray emission appears when the plasma blob/jet is
already far from the central engine, since $\gamma\gamma$ absorption on
background photons is negligible and the pair plasma ahead also becomes
too rare due to annihilation and/or spatial expansion.

Generation and annihilation of a pair plasma should thus be an
intrinsic property of all relativistic blob/jet models. However, the
detection of the annihilation line depends on its luminosity and
viewing angle (Skibo et al.\ 1997).  The described model provides
support to models which include two populations of jet particles,
although the exact flow geometry is still unknown.  According to our
model, the annihilation feature should appear before the high
energy radiation producing a visible anticorrelation on a timescale of
$\Delta_d$ (eq.~[3]); this matches the observations of PKS
0208--512 where the anticorrelation has been noted (Blom et al.\ 1996).

}



\bsk
\baselineskip = 12pt


{\references \ni REFERENCES
\ssk


\ref Bloemen, H., et al. 1995, A\&A, 293, L1
\ref Blom, J.J., et al. 1995, A\&A, 298, L33
\ref Blom, J.J., et al. 1996, A\&AS, 120C, 507
\ref Dermer, C.D., Schlickeiser, R., Mastichiadis, A. 1992, A\&A, 256, L27
\ref Dermer, C.D., Sturner, S.J., Schlickeiser, R. 1997, ApJS, 109, 103
\ref Hartman, R.C., et al.
      1997, in 4th Compton Symp.\ AIP 410,
      AIP, New York, p.307
\ref Henri, G., Pelletier, G., Roland, J. 1993, ApJ, 404, L41
\ref Kanbach, G. 1996, in Workshop on Gamma-ray emitting AGN,
      MPI H-V37-1996, Heidelberg, p.1
\ref Maraschi, L., Ghisellini, G., Celotti, A. 1992, ApJ, 397, L5
\ref Mannheim, K. 1993, A\&A, 269, 67
\ref Mastichiadis, A., Protheroe, R.J. 1990, MNRAS, 246, 279
\ref von Montigny, C., et al. 1995, ApJ, 440, 525
\ref Moskalenko, I.V., Jourdain, E. 1997, A\&A, 325, 401
\ref Roland, J., Hermsen, W. 1995, A\&A, 297, L9
\ref Skibo, J.G., Dermer, C.D., Schlickeiser R. 1997, ApJ, 483, 56
}                      

\end{document}